\documentstyle [12pt, epsfig]{article}
\begin{document}
\hfill {WM-01-115}
\parindent 0.2in

\hfill {\today}
\vskip 0.2in  
\baselineskip 24pt
{
\Large
   \bigskip
   \centerline{Extra neutral gauge bosons and Higgs bosons in an $E_6$-based model}
}
\vskip .2in
\def\bar{\overline}
\centerline{Shuquan Nie\footnote{Email address: sxnie@physics.wm.edu} and Marc Sher\footnote{Email address: sher@physics.wm.edu}}
\bigskip
\centerline {\it Nuclear and Particle Theory Group}
\centerline {\it Physics
Department}
\centerline {\it College of William and Mary, Williamsburg, VA 23187, USA}
\vskip 0.3in
{\narrower\narrower 
Extra neutral gauge bosons and Higgs bosons in an effective low-energy 
$SU(2)_L \times SU(2)_I \times U(1)_Y \times U(1)_{Y^{\prime}}$ model,
which is a subgroup of $E_6$, are studied.  
$SU(2)_I$ is a subgroup of $SU(3)_R$ and commutes with the electric
charge operator, so the three corresponding gauge bosons are neutral.   
Electroweak precision experiments are used to put constraints on  
masses of the extra neutral gauge
bosons and on the mixings between them and the ordinary Z boson,
including constraints arising from a proposed measurment of the weak charge of the proton at Jefferson Lab.
Bounds on and relationships of masses of Higgs bosons in the supersymmetric 
version of the model are also discussed. 
}

\newpage
\section {Introduction}
\hspace{0.2 in}The mass of the Higgs boson of the Standard Model(SM) 
is still undertemined,
although there are recent reports indicating the observation of signals 
at LEP [1-3]. 
The requirement that the vacuum is stable and the perturbation is valid up to 
a large scale(for example, grand unification scale) can bound the mass(es) of Higgs boson(s) \cite{sher}. 
Extra Higgs bosons and
gauge bosons will appear naturally in many extensions of the SM. 
Generally the masses of extra gauge bosons remain unpredicted and may or 
may not be of the order of the electroweak scale. The 
closeness of the observed W and Z boson properties
with the predictions of the SM do not yield  any direct information about 
the masses of extra gauge bosons, but seems to imply that the mixings of W or Z 
with extra gauge bosons should be very small.  

$E_6$ models have been studied widely \cite{Hewett}. 
The maximal subgroup decomposition of $E_6$ 
containing QCD as an explicit factor is $SU(3)_C \times SU(3)_L \times SU(3)_R$, 
from which an effective low energy model $SU(2)_L \times SU(2)_I \times U(1)_Y
\times U(1)_{Y^{\prime}}$ can arise \cite{London}. $SU(2)_I$
commutes with the electric charge operator and the corresponding
gauge bosons are neutral. The most extensive works on the phenomenology of 
this model focused on the 
production of the $W_I$'s in hadron-hadron, $e^+e^-$, and 
$ep$ colliders \cite{Rizzo, Godfrey2}.  
The t-channel production of exotic fermions in the model 
has recently been considered in Ref. \cite{niesher}. In this paper we will study 
the gauge boson and Higgs boson sectors of the model, and bounds on the masses and mixings 
of extra neutral gauge bosons and Higgs bosons will be found.

In Ref. \cite{Abe}, a direct search for extra gauge bosons was reported and lower 
mass limits of approximately $500 \sim 700$ GeV were set, depending on the 
$Z^{\prime}$ couplings. The discovery potential and diagnostic abilities of proposed 
future colliders for new neutral or 
charged gauge bosons were summarized in Ref. \cite{Godfrey}. Even though there is 
as yet no direct experimental evidence of extra gauge bosons,
stringent indirect constraints can be 
put on the mixings and the masses of extra gauge bosons by electroweak 
precision data. In Ref. [12-14], 
such constraints were derived in the $SU(2)_L \times U(1)_Y \times U(1)_{Y^{\prime}}$ model. 
The lower mass limits were generally several hundreds of GeV and were 
competitive with experimental bounds from direct searches.
A good summary of $Z^{\prime}$ searches can be found in Ref. \cite{Leike} and references therein.

The paper is organized as follows. In the next section, 
the model will be described briefly, and
a specific Higgs field assignment to break
$SU(2)_L \times U(1)_Y \times SU(2)_I \times U(1)_{Y^{\prime}}$ 
into $U(1)_{em}$ will be introduced.
Sec. 3 deals with the extra neutral gauge bosons.  
The mixing among neutral gauge bosons will be discussed. 
In Sec. 4, electroweak precision experiments, including Z-pole experiments,
$m_W$ measurements and low-energy neutral current(LENC) experiments will be 
presented, with special attention being paid to a proposed measurement of the 
weak charge of the proton at Jefferson Lab.
In Sec. 5, constraints on the masses of extra neutral gauge bosons and mixings will be 
found. 
In Sec. 6, bounds on and relationships of masses of Higgs bosons appearing in 
the supersymmetric version of the model will be derived. 
Sec. 7 contains our conclusions.
Mass-squared matrices of neutral gauge bosons and Higgs bosons in the model are
given in the Appendices.
 
\section{The Model}
\hspace{0.2 in}There are many phenomenologically acceptable low energy models
which can arise from $E_6$:
\begin{eqnarray} 
(a) &E_6& \rightarrow SU(3)_C \times SU(2)_L \times U(1)_Y \times U(1)_{\eta},  \nonumber \\ 
(b) &E_6& \rightarrow SO(10) \times U(1)_{\psi} \rightarrow SU(5) \times U(1)_{\chi}\times U(1)_{\psi}, \nonumber  \\ 
(c) &E_6& \rightarrow SU(3)_C \times SU(2)_L \times SU(2)_R \times U(1)_L \times U(1)_R,  \nonumber \\ 
(c^{\prime}) &E_6& \rightarrow SU(3)_C \times SU(2)_L \times SU(2)_I \times U(1)_Y \times U(1)_{Y^{\prime}},   
\end{eqnarray} 
where there is only one extra Z, generally called $Z_{\eta}$, 
in model (a). $U(1)_{\psi}$ and $U(1)_{\chi}$ can be combined into
$U(1)_{\theta}$ as $Z^{\prime}(\theta)=Z_{\psi} \cos \theta - Z_{\chi} \sin \theta$ in model (b), 
reducing it to the effective  rank-5 model
$SU(3)_C \times SU(2)_L \times U(1)_Y \times U(1)_{\theta}$, which is  most often considered
in the literature. In particular, $U(1)_{\eta}$ corresponds to $\theta = \arcsin \sqrt{3/8}$.
Model (c) and $(c^{\prime})$ come from the subgroup $SU(3)_C \times
SU(3)_L \times SU(3)_R$. The {\bf 27}-dimensional fundamental representation of $E_6$ has the
branching rule
\begin{equation} {\bf 27}=\underbrace{({\bf 3}^c,{\bf 3},{\bf 1})}_{q}+\underbrace{
(\bar{{\bf 3}}^c, {\bf 1}, \bar{{\bf 3}})}_{\bar{q}} +\underbrace{({\bf 1}^c, \bar{{\bf
3}}, {\bf 3})}_{l} \ ,
\end{equation} and the particles of the first family are assigned as
\begin{eqnarray}
  \left( \begin{array}{c} u\\d\\h \end{array} \right)+
  \left( \begin{array}{lcr} u^c & d^c & h^c \end{array} \right)+
  \left( \begin{array}{ccc} E^c & \nu & N \\ N^c & e & E\\e^c & \nu^c & S^c
  \end{array} \right),  
\end{eqnarray}
where $SU(3)_L$ operates vertically and $SU(3)_R$ operates
horizontally. (Different symbols for these particles may be used in the literature.)

The most common pattern of breaking the $SU(3)_R$ factor is to break the {\bf 3} of
$SU(3)_R$ into {\bf 2}+{\bf 1}, so that $(u^c, d^c)$ forms an $SU(2)_R$ doublet with
$h^c$ as a $SU(2)_R$ singlet. This gives model (c), the familiar left-right  symmetric
model \cite{Candelas}. Model (c) can be reduced further to an effective rank-5  model
with $U(1)_{V=L+R}$. Another possibility, resulting in model $(c^{\prime})$, is to
break the {\bf 3} of the $SU(3)_R$ into ${\bf 1} + {\bf 2}$ so  that $(d^c, h^c)$ forms
an $SU(2)$ doublet with $u^c$ as a singlet. In this option, the $SU(2)$ doesn't
contribute to the electromagnetic charge operator and it is called $SU(2)_I$ (I stands
for Inert). Then the vector gauge bosons corresponding to $SU(2)_I$ are neutral.
 
At the $SU(2)_L \times SU(2)_I \times U(1)_Y
\times U(1)_{Y^{\prime}}$ level, a single generation of fermions can be represented as
\begin{eqnarray}
\left( \begin{array}{cc} \nu & N\\ e^{-} & E^{-}  \end{array} \right)_L, \hspace{2mm}
   \left( \begin{array}{c} u \\ d \end{array} \right)_L, \hspace{2mm}
   \left( \begin{array}{lr} d^c & h^c \end{array} \right)_L, \hspace{2mm}
   \left( \begin{array}{c} E^c \\ N^c \end{array} \right)_L, \hspace{2mm}
   \left( \begin{array}{lr} \nu^c & S^c \end{array} \right)_L, \hspace{2mm}
   h_L, \hspace{2mm} e_L^c, \hspace{2mm} u^c_L,
\end{eqnarray}

\begin{center}
Table {\bf 1} The quantum numbers of fermions in {\bf 27} of $E_6$ \\ 
at the $SU(2)_L \times SU(2)_I \times U(1)_Y \times U(1)_{Y^{\prime}}$ level.\\ 
\vspace{0.5 cm}
\begin{tabular}{c|c|c|c|c|c} \hline \hline
State& $T_{3L}$  & $T_{3I}$ &  $Y$ & $Y^{\prime}$ & $Q_{em}=T_{3L}+Y/2$ \\ \hline
$u$     & 1/2  &  0  &1/3 &2/3  & 2/3 \\ \hline
$d$     & -1/2 &  0  &1/3 &2/3  & -1/3 \\ \hline
$u^c$   & 0    &  0  &-4/3& 2/3 & -2/3 \\ \hline
$d^c$   & 0    & 1/2 & 2/3& -1/3& 1/3 \\  \hline
$h$     & 0    & 0  & -2/3&-4/3 & -1/3 \\ \hline
$h^c$   & 0    &-1/2& 2/3&-1/3 & 1/3 \\ \hline
$e^{-}$ &-1/2  &1/2 & -1 & -1/3& -1 \\ \hline
$e^{c}$ & 0    & 0 & 2  & 2/3 & 1 \\ \hline
$E^{-}$ & -1/2 &-1/2& -1 &-1/3 & -1 \\ \hline
$E^{c}$ & 1/2  & 0  & 1  &-4/3 & 1 \\ \hline
$\nu$&1/2 & 1/2& -1 & -1/3& 0 \\ \hline
$\nu ^c$&0 & 1/2&0   & 5/3 & 0 \\ \hline
$N$        &1/2&-1/2&-1  & -1/3& 0 \\ \hline
$N^c$      &-1/2& 0  & 1  & -4/3& 0 \\ \hline
$S^c$      & 0 &-1/2& 0  & 5/3 & 0 \\ \hline \hline
\end{tabular}
\end{center}

\noindent where $SU(2)_{L(I)}$ acts vertically (horizontally). The quantum numbers of particles 
are listed in Table {\bf 1}.

In Ref. \cite{Haber} the Higgs structure necessary to break $SU(2)_L \times SU(2)_I 
\times U(1)_Y \times U(1)_{Y^{\prime}}$ down to $U(1)_{em}$ was discussed.
The Higgs multiplets are 
\begin{eqnarray}
{\sl H_2} \equiv \left( \begin{array}{c} H_2^{+} \\ H_2^{0} \end{array} 
\right),\hspace{2mm}
{\cal H} \equiv \left( \begin{array}{c c} H_1^{0} \hspace{4mm} \tilde{{\nu}}  \\ 
H_1^{-} \hspace{4mm} \tilde{e^{-}} \end{array} \right),\hspace{2mm}
 {\sl N} \equiv \left( \begin{array}{l r} N_2 \hspace{4 mm} N_1 \end{array} \right),
\hspace{2mm}
{\sl N^{\prime}} \equiv \left( \begin{array}{l r } N_2^{\prime} \hspace{4 mm} N_1^{\prime}
\end{array} \right),  
\end{eqnarray}
with $SU(2)_L$ acting in the vertical direction and $SU(2)_I$ acting in 
the horizational direction. The U(1) quantum numbers are:
$Y({\sl H_2})=1,Y({\cal H})=-1, Y({\sl N})=Y({\sl N^{\prime}})=0$, and $
Y^{\prime}({\sl H_2})=4/3,Y^{\prime}({\cal H})=1/3, Y^{\prime}({\sl N})
=Y^{\prime}({\sl N^{\prime}})=-5/3$. The doublets {\sl N} and ${\sl N^{\prime}}$
are also neutral. Note that two N doublets are needed. The reason can be seen
in the limit where the model is broken down to the SM at a scale much greater than
the electroweak scale. A single N doublet can only break $SU(2)_I \times U(1)_{Y^{\prime}}$
down to U(1), leaving an extra unbroken U(1) symmetry. 

The multiplets can get vacuum expectation values in the following way,
\begin{equation}
\langle{\sl H_2} \rangle = \left( \begin{array}{c} 0 \\ v_2
\end{array} 
\right),\hspace{2mm}
\langle {\cal H} \rangle = \left( \begin{array}{c c} v_1 \hspace{4mm} 
v_3 \\ 
0 \hspace{4mm} 0 \end{array} \right),\hspace{2mm}
\langle {\sl N} \rangle = \left( \begin{array}{l r} n_2 \hspace{4 mm} n_1 \end{array} \right),
\hspace{2mm}
\langle {\sl N^{\prime}} \rangle =  \left( \begin{array}{l r} n_2^{\prime} \hspace{4 mm} n_1^{\prime} 
\end{array} \right).  
\end{equation}
Since we are not considering the spontaneous CP violation,  
the phase of the Higgs fields can be chosen such that 
all of vacuum expectation values 
are real and positive. There appear to be
seven vacuum expectation values in the model, but one of them can be set to
zero by performing an $SU(2)_I$ rotation. So there are only six physically
relevant vacuum expectation values.

\section {Extra neutral gauge bosons and mixings}
\hspace{0.2 in}In the $SU(2)_L \times U(1)_Y \times SU(2)_I \times U(1)_{Y^{\prime}}$ model, the neutral gauge fields include 
the ordinary Z coming from  $SU(2)_L \times U(1)_Y$; $W_I^1$, $W_{I}^2$ and $W_{I}^3$ for the $SU(2)_I$ group and B for $U(1)_{Y^{\prime}}$. 
(We will use linear combinations $W_I^{\pm}=(W_I^1 \mp i W_I^2)/ \sqrt{2}$ instead of $W_I^1$ and $W_I^2$, here $\pm$ is just a 
convention as they are neutral.) After the spontaneous 
symmetry breaking mechanism described in the previous section, the mass-squared matrix for the neutral gauge bosons 
is a symmetric $5 \times 5 $ matrix, whose elements are listed in the appendices.

It is apparent that there are mixings among the neutral gauge bosons. It is impossible to diagonalize the matrix analytically. 
Numerical calculations must be needed to get the eigenstates and corresponding eigenvalues. 

It is noted that the 
elements in the first row(column) are independent of the vacuum expectation values $n_i$ and $n_i^{\prime}$(i=1,2). Therefore when they are very large, the mixing should be small. In this 
decoupling limit, the only observable neutral gauge boson is the ordinary Z and its mass should be the exact value measured experimentally. The extra neutral 
gauge bosons are not yet accessible experimentally, 
but their existence will have effects in electroweak radiative corrections.
 
In order to find mass eigenstates and mixing angles, the mass-squared matrix ${\cal M}^2$ can be split into two parts
\begin{eqnarray}
{\cal M}^2&=& {\cal M}_1^2 +{\cal M}_2^2  \nonumber \\
        &=& \left(
\begin{array} {ccccc}
m_Z^2 & 0 & 0 & 0 & 0 \\
0 & m_{W_I^3}^2 & m_{23} & m_{24} & m_{25} \\
0 & m_{23} & m_{B}^2 & m_{34} & m_{35} \\ 
0 & m_{24} & m_{34} &    0   & m_{W_I^{\pm}}^2 \\
0 & m_{25} & m_{35} & m_{W_I^{\pm}}^2 & 0 
\end{array}
\right) + \left(
\begin{array} {ccccc}
0 & m_{12} & m_{13} & m_{14} & m_{15} \\
m_{12} & 0 & 0 & 0 & 0 \\
m_{13} & 0 & 0 & 0 & 0 \\ 
m_{14} & 0 & 0 & 0 & 0 \\
m_{15} & 0 & 0 & 0 & 0 
\end{array}
\right)_{\ .} 
\end{eqnarray}
First we can use a $5 \times 5 $ unitary matrix ${\bf U}_1$ to diagonalize ${\cal M}_1^2$, 
and ${\bf U}_1$ can have  the form 
\begin{equation}
{\bf U}_1=\left(
\begin{array}{cc}
1 & 0 \\
0 & {\bf u}_1
\end{array}
\right)_{\ ,} 
\end{equation}
where ${\bf u}_1$ is a $4 \times 4$ unitary matrix.
This is to find mass eigenstates of extra neutral gauge bosons.
There is no mixing of ordinary Z boson with extra neutral gauge bosons at this 
stage.                                          
Then the total mass-squared matrix for the neutral gauge bosons under the new basis has the form 
\begin{eqnarray}
{\cal M}^{\prime 2} &=& {\cal M}_1^{\prime 2} +{\cal M}_2^{\prime 2} \nonumber \\
                  &=& \left( \begin{array}{ccccc}
                             m_Z^2 & m_{12}^{\prime} & m_{13}^{\prime} & m_{14}^{\prime} &  m_{15}^{\prime} \\
                             m_{12}^{\prime} & m_{Z_2}^2 & & & \\
                             m_{13}^{\prime} & & m_{Z_3}^2 & & \\ 
                             m_{14}^{\prime} & & & m_{Z_4}^2 & \\
                             m_{15}^{\prime} & & & & m_{Z_5}^2 
                             \end{array}
                        \right)_{\ .}
\end{eqnarray}
${\cal M}^{\prime 2}$ can be principally diagonalized by another unitary matrix ${\bf U}_2$, 
then we can get a unitary matrix ${\bf U}= {\bf U}_2 \times {\bf U}_1$ 
which can be used to diagonalize the original matrix ${\cal M}^2$.  
The mixings of ordinary Z boson with extra neutral 
gauge bosons occur in this transformation. 
For small mixings, the elements of ${\bf U}_2$ will have the 
following properties
\begin{eqnarray}
({\bf U}_2)_{11} &\sim& 1.0, \nonumber \\
({\bf U}_2)_{j1} &\sim& \left( \frac{m_Z^2-m_{Z_1}^2}{m_{Z_j^{\prime}}^2-m_Z^2} \right)^{1/2}, \nonumber \\
({\bf U}_2)_{jk} &\sim& 0,\hspace{1 cm}  j \neq k. 
\end{eqnarray} 
Therefore $({\bf U}_2)_{j1}$ can be treated as effective mixing angles.
 
The couplings between neutral gauge bosons and fermions, which will give neutral current processes, are 
\begin{eqnarray}
{\cal L}_{NC}&=&-\sum_{f, \alpha} \{ g_Z \bar{f}_{\alpha} \gamma^{\mu} \left( T_{3L}^{f_{\alpha}}-Q_{f_{\alpha}} \sin^2 \theta_W \right) f_{\alpha} Z_{\mu}
+g_{Y^{\prime}} Y^{\prime}_{f_{\alpha}}/2 \bar{f}_{\alpha} \gamma^{\mu} f_{\alpha} B_{\mu} \nonumber \\
 & & +g_I T_{3I}^{f_{\alpha}} \bar{f}_{\alpha} \gamma^{\mu} f_{\alpha} W^3_{I \mu} \}, 
\end{eqnarray}
where the first term in the brackets represents the SM neutral currents, the second and 
third terms represent additional neutral currents introduced by extra neutral gauge bosons, 
and $g_Z=g_L/\cos \theta_W = g_Y / \sin \theta_W$. The symbol $f_{\alpha}$ denotes the leptons or quarks with the chirality $\alpha$ ($\alpha=L$ or $R$). The quantum numbers
$T_{3L}^{f_{\alpha}}$, $Q_{f_{\alpha}}$, $Y^{\prime}_{f_{\alpha}}$ and $T_{3I}^{f_{\alpha}}$ can be read from Table {\bf 1}. The flavor-changing neutral currents caused by $W_I^{\pm}$
involve heavy fermions and will not be included here. 

After the ${\bf U}_1$-transformation, the interaction Lagrangian changes as 
\begin{eqnarray}
{\cal L}_{NC}&=&-\sum_{f, \alpha} \{ g_Z \bar{f}_{\alpha} \gamma^{\mu} \left( T_{3L}^{f_{\alpha}}-Q_{f_{\alpha}} \sin^2 \theta_W \right) f_{\alpha} Z_{\mu} \nonumber \\
 & & +g_{Y^{\prime}} Y^{\prime}_{f_{\alpha}}/2 \bar{f}_{\alpha} \gamma^{\mu} f_{\alpha} \sum_{j \neq 1}({\bf U}_1)_{3j}Z_{j \mu} \nonumber \\
 & & +g_I T_{3I}^{f_{\alpha}} \bar{f}_{\alpha} \gamma^{\mu} f_{\alpha} \sum_{j \neq 1}({\bf U}_1)_{2j}Z_{j \mu}  \}. 
\end{eqnarray}
where the first term is unchanged because there is no mixing of ordinary Z boson with extra neutral gauge bosons. 
Considering the ${\bf U}_2$-transformation, the final interaction Lagrangian is given as
\begin{eqnarray}
{\cal L}_{NC}&=&-\sum_{f, \alpha} \{ g_Z \bar{f}_{\alpha} \gamma^{\mu} \left( T_{3L}^{f_{\alpha}}-Q_{f_{\alpha}} \sin^2 \theta_W \right) f_{\alpha} [({\bf U}_2)_{11} Z_{1\mu} + \sum_{j \neq 1}({\bf U}_2)_{1j} Z_{j\mu}^{\prime}] \nonumber \\
 & & +g_{Y^{\prime}} Y^{\prime}_{f_{\alpha}}/2 \bar{f}_{\alpha} \gamma^{\mu} f_{\alpha}  \sum _{j \neq 1} ({\bf U}_1)_{3j}[({\bf U}_2)_{j1} Z_{1\mu} + \sum_{k \neq 1}({\bf U}_2)_{jk} Z_{k\mu}^{\prime}] \nonumber \\
 & & +g_I T_{3I}^{f_{\alpha}} \bar{f}_{\alpha} \gamma^{\mu} f_{\alpha} \sum _{j \neq 1} ({\bf U}_1)_{2j} [({\bf U}_2)_{j1} Z_{1\mu} + \sum_{k \neq 1}({\bf U}_2)_{jk} Z_{k\mu}^{\prime}]  \}, 
\end{eqnarray}
The contributions from the term $({\bf U}_2)_{1j} Z_{j\mu}^{\prime}$ 
can be omitted in our analysis because they are combinations of mixings and exchanges of extra neutral gauge bosons and should be very small.    

Due to the mixings, the mass, $m_{Z_1}$ of the observed Z boson is shifted from the SM prediction $m_Z$. 
\begin{equation}
\Delta m^2 \equiv m^2_{Z_1}-m_Z^2 \leq 0.
\end{equation}
The presence of this mass shift will affect the T-parameter \cite{Peskin} at tree level.  
From Ref. \cite{Cho1}, the T-parameter is expressed in terms of
the effective form factors $\bar{g}^2_{Z}(0)$, $ \bar{g}^2_{W}(0)$ and the fine structure constant $\alpha$ as 
\begin{eqnarray}
\alpha T &\equiv& 1- \frac{\bar{g}^2_{W}(0)}{m_W^2}\frac{ m_{Z_1}^2}{\bar{g}^2_{Z}(0)} \nonumber \\
&=& \alpha ( T_{SM} + T_{new} ),
\end{eqnarray}
where $T_{SM}$ and the new physics contribution $T_{new}$ are given by
\begin{eqnarray}
\alpha T_{SM} &=& 1- \frac{\bar{g}^2_{W}(0)}{m_W^2}\frac{ m_{Z}^2}{\bar{g}^2_{Z}(0)},  \\
\alpha T_{new}&=& - \frac{\Delta m^2}{m_{Z_1}^2} \geq 0.
\end{eqnarray} 
It is noted that the positiveness of $T_{new}$ is attributed to the mixings 
which always lower the mass of the ordinary Z boson.
The effects of Z-$Z^{\prime}$ mixings can be 
described by the effective mixing angles and the positive $T_{new}$. 

\section {Electroweak observables }
\hspace{0.2 in}The experimental data used to put indirect constraints on extra neutral gauge bosons 
are summarized in Table {\bf 2}. The data includes the Z-pole 
experiments, the W boson mass measurement and LENC experiments.
They are updated from Ref. \cite{PDG, langacker}. The family universality is assumed in our analysis.

\newpage

\begin{center}
Table {\bf 2} Summary of precision electroweak measurements used in our analysis. \\
\vspace{0.5 cm}
\begin{tabular}{|c|c|} \hline \hline
\multicolumn{2}{|c|}{{\bf Z-pole experiments}} \\ \hline
$m_Z$ (GeV)     &  91.1872 $\pm$ 0.0021 \\ \hline
$\Gamma_Z$ (GeV)  &  2.4944  $\pm$ 0.0024 \\ \hline
$\sigma^{0}_{h}$ (nb) & 41.544 $\pm$ 0.037 \\ \hline
$R_l$ & 20.784 $\pm$ 0.023 \\ \hline
$A^{0,l}_{FB}$ & 0.0170 $\pm$ 0.0009 \\ \hline
$A_{\tau}$ & 0.1425 $\pm$ 0.0043 \\ \hline
$A_e$ & 0.1511 $\pm$ 0.0019 \\ \hline
$R_b$ & 0.21642 $\pm$ 0.00073 \\ \hline
$R_c$ & 0.1674 $\pm$ 0.0038 \\ \hline
$A^{0,b}_{FB}$ & 0.0988 $\pm$ 0.0020 \\ \hline
$A_{FB}^{0,c}$ & 0.0692 $\pm$ 0.0037 \\ \hline
$A^0_{LR} $ & 0.1495 $\pm$ 0.0017 \\ \hline
$A_b$ & 0.911 $\pm$ 0.025 \\ \hline
$A_c$ & 0.630 $\pm$ 0.026 \\ \hline
\multicolumn{2}{|c|} {{\bf W-mass measurement}} \\ \hline
$m_W$ (GeV) & 80.394 $\pm$ 0.042 \\ \hline 
\multicolumn{2}{|c|} {{\bf LENC experiments}} \\ \hline
$A_{SLAC}$ & 0.80 $\pm$ 0.058 \\ \hline
$A_{CERN}$ & -1.57 $\pm$ 0.38 \\ \hline
$A_{Bates}$ & -0.137 $\pm$ 0.033 \\ \hline
$A_{Mainz}$ & -0.94 $\pm$ 0.19 \\ \hline
$Q_W (^{133}_{55}\mathrm{Cs})$ & -72.06 $\pm$ 0.44 \\ \hline
$K_{FH}$ & 0.3247 $\pm$ 0.0040 \\ \hline
$K_{CCFR} $ & 0.5820 $\pm$ 0.0049 \\ \hline
$g_{LL}^{\nu_{\mu} e}$ & -0.269 $\pm$ 0.011 \\ \hline
$g_{LR}^{\nu_{\mu} e}$ & 0.234 $\pm$ 0.011 \\ \hline \hline    
\end{tabular}
\end{center}

In addition to the electroweak observables generally used in the literature, we also consider the possible constraint 
arising from the weak charge of the proton, which is proposed to be measured at Jefferson Lab.
In contrast to the weak charge of a heavy atom, the weak charge
of the proton is fortuitously suppressed in the SM. Therefore it  
is very sensitive to the contributions from new physics. Additionally it is twice as sensitive to new u-quark interactions as it is 
to new d-quark physics. In the model considered here the right-handed u-quark and d-quark have different isospin contents under $SU(2)_I$, 
so it is advantageous to consider the constraints arising from the anticipated measurement. 
The theoretical prediction \cite{Cho1} for the weak charge of the proton can be derived 
\begin{eqnarray}
Q_W^P&=& 0.07202-0.01362 \Delta S + 0.00954 \Delta T +2(2 \Delta C_{1u}+\Delta C_{1d}). 
\end{eqnarray} 

\section{Constraints on extra neutral gauge bosons}
\hspace{0.2 in}Using the electroweak precision data, constraints on mixing angles and masses of extra neutral gauge bosons 
can be obtained from the standard $\chi^2$ analysis.  
For simplicity, $S_{new}$ and $U_{new}$
will be set zero because they are very small.     
Through our analysis, we will use precisely determined parameters
$m_{Z_1}$, $G_f$ and $\bar{\alpha}(m^2_{Z_1})$
as inputs.
The Higgs mass dependence of 
the results are ignored for simplicity. 
We set the top quark mass $m_t=175 \ \mathrm{GeV}$ and Higgs boson mass $m_H=100 \ \mathrm{GeV}$ in our analysis.
We first obtain the constraints from 
Z-pole experiments and $m_W$ measurement only, and then we combine the LENC experiments with them to get further constraints.    
Finally we will study the possible constraints which would arise from measuring the weak charge of the proton. 

\subsection{Constraints from Z-pole and $m_W$ data}
\begin{figure}
\centerline{ \epsfysize 4in \rotatebox{270}{\epsfbox{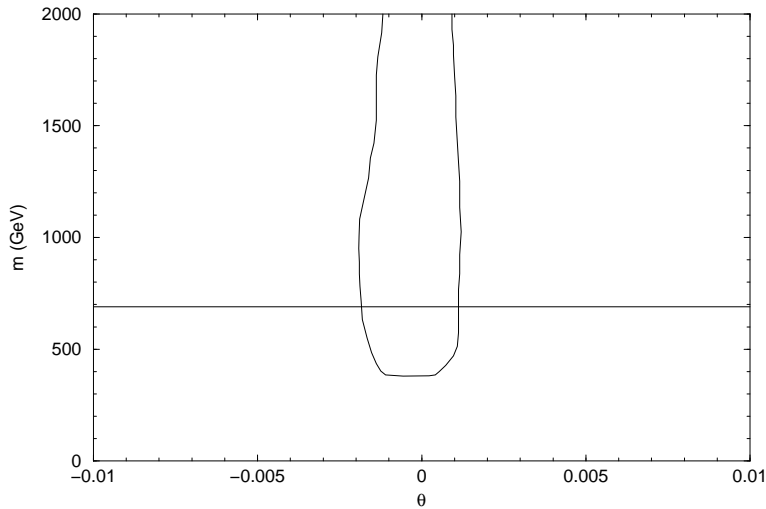}}  }
\caption{The contour of $ \Delta \chi ^2=\chi^2 -\chi^2_{SM}=1.0$ for the lightest extra 
neutral gauge boson. The constraint is obtained 
by use of Z-pole experiments and $m_W$ measurement. As a reference the 
lower direct production limit
from CDF \cite{Abe} for the sequential $Z_{SM}$ is also shown.}
\protect \label{figzw}
\end{figure}

\hspace{0.2 in}From the previous analysis, it is found that the Z-pole experiments are related to mixings and the T-parameter, while $m_W$ is only 
relevant for the T-parameter. 
If we set all mixing angles and $T_{new}$ equal to zero, it will 
give the fit for the SM. 
It serves as a reference because the SM
fits the experiments very well. 
Defining $\Delta \chi^2=\chi^2-\chi^2_{SM}$, by requiring acceptable $\Delta \chi^2$ we can get constraints on the mixings and the masses of extra neutral gauge bosons. 
The result for $\Delta \chi^2=1.0$ is illustrated in Fig. \ref{figzw}. The lower mass limit for the lightest extra gauge boson 
is about $400$ GeV. It seems that the model allows for the existence of a comparatively light
extra neutral gauge boson. But we will find in the following that this is not true when LENC 
experiments are included. The mixing angles are found to be 
very small, namely $|\theta| \leq 0.003$. 

The sequential $Z_{SM}$ boson \cite{Barger} is defined to have the same couplings to fermions as the SM Z boson.
Such a boson is not expected in the context of gauge theories unless it has different couplings to exotic fermions than the 
ordinary Z. However, it serves as a useful reference case when comparing constraints from various sources. The direct production
limit for the sequential $Z_{SM}$ boson from Ref. \cite{Abe} is about $690$ GeV. It is assumed that all exotic decay channels are forbidden, and the bound has to be relaxed by about 
$100$ to $150$ GeV when all exotic decays(including channels involving superparticles) are kinetically allowed. It is found 
that, at this time, the lower mass limit for the lightest extra neutral gauge boson is much lower than the direct production limit for 
the sequential $Z_{SM}$ boson.    

\subsection{Constrains from Z-pole + $m_W$ + LENC data}
\begin{figure}
\centerline{ \epsfysize 4in \rotatebox{270}{\epsfbox{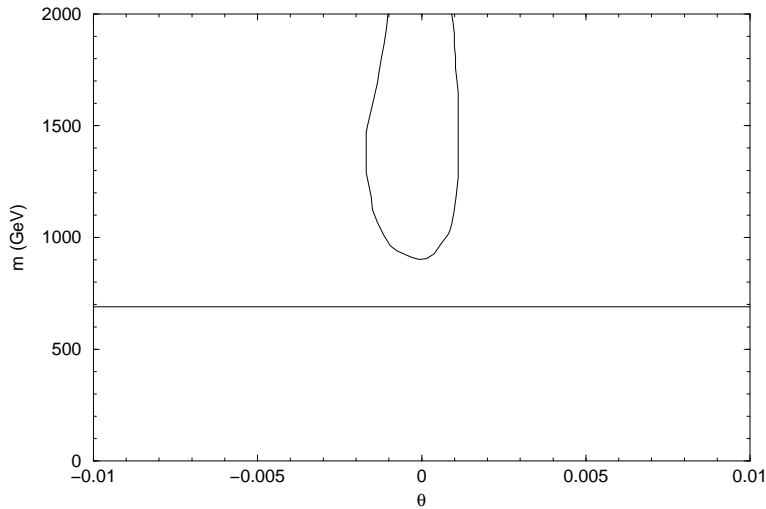}}  }
\caption{The contour of $\Delta \chi ^2=\chi^2 -\chi^2_{SM}=1.0$ for the lightest extra 
neutral gauge boson. The constraint is obtained 
by use of Z-pole experiments, $m_W$ measurement and LENC experiments. As a reference the 
lower direct production limit
from CDF \cite{Abe} for the sequential $Z_{SM}$ is also shown.}
\protect \label{figzwl}
\end{figure}

\hspace{0.2 in}The LENC experiments can get contributions from the exchanges of extra neutral gauge bosons, which can be approximated by 
contact interactions. The contact interactions are inversely-proportional to the masses of the extra gauge bosons exchanged in 
the processes. So the LENC experiments can put stringent constraints on the masses of extra neutral gauge bosons. The results of 
fitting Z-pole experiments, $m_W$ measurement and LENC experiments are
shown in Fig. \ref{figzwl}. The lower mass limits for the extra neutral gauge bosons are raised much higher
than those without LENC experiments. The lower mass bound for the lightest extra gauge boson is about $900$ GeV. 
It is higher than the direct production limit for the sequential $Z_{SM}$ boson.  

In Ref. \cite{Erler2}, similar constraints on various possible extra $Z^{\prime}$ bosons were studied.
In all cases the mixing angles are severely constrained($\sin \theta < 0.01$), and the lower mass limit 
are generally of the order of several hundred GeV, depending on the specific models considered. 

In the model considered here, from the appendices, $m_{W_I^3}^2 \sim m_B^2$ assuming that $g_I=g_L$ and $g_{Y^{\prime}}=g_Y$. 
It is apparent that $m_{W^{\pm}_I}$ is degenerate with $m_{W_I^3}$ without mixing. Generally the lightest
extra neutral gauge boson mainly consists of $W_I^3$, or $Z_I$. 
It is noted that $Z_I$ corresponds to $Z^{\prime}(\theta=-\arcsin \sqrt{5/8})$ and is orthogonal to $Z_{\eta}$. 
There is no mass limit on $Z_I$ from electroweak precision data available in the literature. 
From constraints on $Z_{\psi}$, $Z_{\chi}$ and $Z_{\eta}$ \cite{Erler2}, it could be inferred that the mass limit on $Z_I$ 
would be about $430$ GeV at $95 \%$ CL.  
In Ref. \cite{Abe} the lower mass limit of $565$ GeV for $Z_I$ was set by direct search for heavy neutral gauge bosons
with the Collider Detector at Fermilab.
Our mass limit on the lightest extra neutral gauge boson is much higher mainly 
due to more updateded data used in our analysis.

It should be pointed out that an updated value for $Q_W(\mathrm{Cs})=-72.06(28)_{expt}(34)_{theor}$ has been reported \cite{BennettWieman}.
The experimental precision was improved and indicated a 2.5$\sigma$ deviation from the prediction of the SM.
The possibility that the discrepancy is due to contributions from new physics has been suggested. 
In Ref. \cite{Rosner, Casalbuoni} it was shown that the contribution from the exchange 
of an extra U(1) boson could explain the data without $Z-Z^{\prime}$ mixing. Some models which would give negative contibutions to 
$Q_W(\mathrm{Cs})$, such as $Z_{SM}$ and $Z_{\eta}$, were excluded at $99\%$ CL. The existance of $Z_I$ with a central value of about 
$760 \ \mathrm{GeV}$ could explain the deviation. 

Of cousre, a 2.5$\sigma$ discrepance is insufficient to claim a discovery, so we have used the data to determine lower mass bounds
and mixings of additional neutral gauge bosons. 
It put much stronger constraints on the mass and mixing of the lightest extra neutral gauge boson than the old data.
 
From Ref. \cite{Godfrey} the typical bounds achievable on extra neutral or charged gauge bosons $m_{Z^{\prime}(W^{\prime})}$ at the 
coming colliders such as Tevatron, LHC, 500 GeV NLC and 1 TeV NLC are approximately 1 TeV, 4 TeV, 1-3 TeV and 
2-6 TeV correspondingly. Therefore the extra neutral gauge bosons in the model could be studied well in the coming colliding 
experiments.

\begin{figure}
\centerline{ \epsfysize 4in \rotatebox{270}{\epsfbox{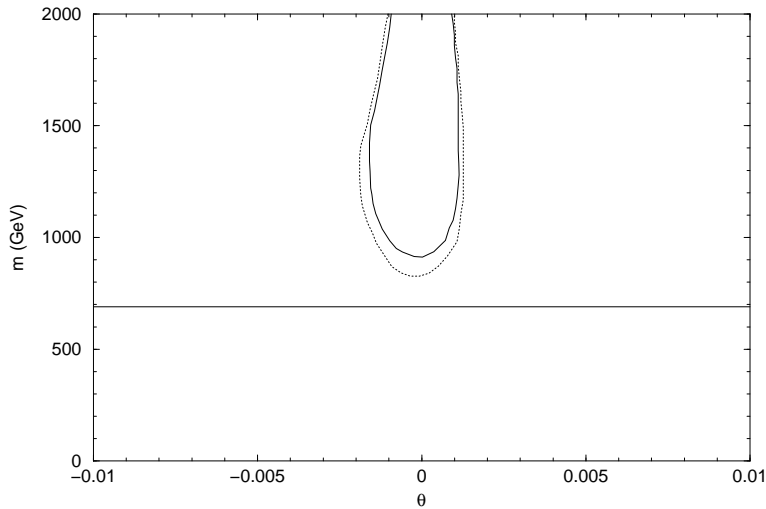}}  }
\caption{The contour of $ \Delta \chi ^2=\chi^2 -\chi^2_{SM}=1.0$ for the lightest extra 
neutral gauge boson with the data of Z-pole experiments, $m_W$ measurement, LENC experiments 
and proposed measurement of the weak charge
of the proton with the precision level of $3\%$. 
As a reference the lower direct production limit
from CDF \cite{Abe} for the sequential $Z_{SM}$ is also shown. 
The contour of $\Delta \chi ^2 =2.0$ is also shown(dotted line).}
\protect \label{figzwlp}
\end{figure}

\subsection{Constrains from Z-pole + $m_W$ + LENC data + $Q_W^P$}
\hspace{0.2 in}In Ref. \cite{Jefferson Lab}, it is proposed to measure the weak charge of the proton, $Q_W^P$, with parity-violating {\it ep} scattering at
$Q^2=0.03 (\mathrm{GeV/c})^2$ at Jefferson Lab. A high statistical accuracy is expected to be achieved with the current facility. Specifically, 
$\Delta Q_W^P / Q_W^{0P} \sim 4 \% $ or better is possible. Fig. \ref{figzwlp} illustrates the constraints on the lightest extra neutral gauge boson
including the $Q_W^P$ assuming that the precision level is $3\%$. It is found that the lower mass bound of the 
the lightest extra neutral gauge boson 
is almost same as the constraint with the data of Z-pole expreiments, $m_W$ and LENC experiments. 
Should the weak charge of the proton be measured with a high precision level,
it would yield competitive constraints on the model. 

In Ref. \cite{Musolf}, the new physics sensitivity of a variety of low energy
parity-violating observables was analyzed. Taken as an example, present and prospective mass limits on an additional gauge boson, $Z_{\chi}$, were given.
Were the precison of measuring the weak charge of the proton $10\%$($3\%$), the lower bound would be $585$($1100$) GeV. This is compatible with our result.

\section {Bounds on Higgs bosons}
\hspace{0.2 in}
There are a large number of Higgs bosons in the model:
6 scalar, 3 pseudoscalar and 4 charged Higgs bosons. In general, 
the scalar potential will have too many parameters to make any meaningful statement
about masses of Higgs bosons. However, in the supersymmetric version of the model, the scalar potential
is highly constrained.
 
The most general superpotential satisfying gauge invariance can be written as
\begin{equation}
W=\lambda {\sl H_2} {\cal H} {\sl N} + \lambda^{\prime} {\sl H_2} {\cal H} 
{\sl N^{\prime}}.
\end{equation}

Here ${\sl H_2}{\cal H} {\sl  N}$ means $\varepsilon_{ij} {\sl H_2}^i {\cal H}^{\alpha j}
\varepsilon_{\alpha \beta} {\sl N}^{\beta}$, i, j are $SU(2)_L$ indices and 
$\alpha,\beta$ are $SU(2)_I$ indices. The scalar potential is given by 
\begin{equation}
V=V_F+V_D+V_{soft},
\end{equation}
where 
\begin{equation}
V_F=\sum_{i}\mid \partial W / \partial \phi_i |^{2}
\end{equation} is the F-term, the sum runs over all complex scalar
$\phi _i$'s appearing in the theory.
\begin{equation}
V_D=1/2 \sum_{a} |\sum_{i}(g_a \phi_i^{\dag} T^a  \phi_i)+ \xi_a |^{2}
\end{equation} is the D-term, $T^a$ represent generators of corresponding
gauge groups and $g_a$ coupling constants. 
The $\xi$ terms only exist if $a$ labels a U(1) generator, and in our
consideration they are set to zero for simplicity.
\begin{eqnarray}
V_{soft}&=& m_{\cal H}^2 Tr({\cal H^{\dag} H})
+m_{\sl H_2}^2 {\sl H_2^{\dag} H_2}
+m_{\sl N}^2 {\sl N^{\dag}} {\sl N}+m_{\sl N^{\prime}}^2 {\sl N^{\prime \dag}}
{\sl N^{\prime}} \nonumber \\ 
& &- \lambda {\sl A}  ({\sl H_2} {\cal H}{\sl N}+h.c.)
- \lambda^{\prime} {\sl A^{\prime}}({\sl H_2} {\cal H}
{\sl N^{\prime}}+h.c.) - m_3^2 (N^{\dag} N^{\prime}+h.c.) 
\end{eqnarray} 
are soft supersymmetry breaking terms. The soft supersymmetry breaking parameters
will be considered completely arbitrary, therefore we only study the tree-level potential. 
The radiative corrections to the potential will not significantly affect the results because  
the primary effects of the radiative corrections are to change the effective soft supersymmetry
breaking terms. The exception is due to top quark contribution, proportional to $m^4_{top}$, 
and it will increase some mass limits by up to $20$ GeV.

The complete potential has nine parameters: $\lambda$, $\lambda^{\prime}$,
the coefficients of the two trilinear terms ${\sl A}$ and ${\sl A^{\prime}}$,
the four mass-squared parameters $m_{\cal H}^2$, $m_{\sl H_2}^2$,
$m_{\sl N}^2$ and $m_{\sl N^{\prime}}^2$, and $m_3^2$. Six of them can be transferred to vacuum 
expectation values, thus three undetermined parameters remain, which we take 
to be  $\lambda$, $\lambda^{\prime}$ and $m_3^2$. All the parameters are chosen to be real, therefore the 
scalar potential is CP invariant.

It is straightforward but tedious to work out the mass-squared matrices for 
various Higgs bosons, which are given in the appendices. The mass-squared matrices
for the neutral scalars and pseudoscalars are $7 \times 7$ matrices. 
The two matrices are decoupled from each other
because the scalar potential is CP invariant. The former
must have one zero eigenvalue and the latter must have four zero eigenvalues,
corresponding to the five Goldstone bosons eaten by the five massive neutral
vector gauge bosons[the zero eigenvalue of the scalar mass-squared matrix 
corresponds to the freedom to perform an $SU(2)_I$ rotation in order to set
one of neutral vacuum expectation values to zero]. The mass-squared matrices
for charged Higgs scalars are $3 \times 3$ matrices. The positive 
states and negative states decouple, and they share the same mass-squared 
matrix. There is one zero eigenvalue for each of them in order to 
produce masses for two charged vector bosons of $SU(2)_L$. As we must resort to
numerical techniques to find the eigenvalues of the Higgs bosons, the presence
of the required number of zero eigenvalues provides an excellent check on our
numerical calculation. As another check, we found that there exists a
relationship
\begin{equation}
Tr {\sl M_{\phi}}^2 =Tr {\sl M_Z}^2 + Tr {\sl M_{H_3^0}}^2,
\end{equation}      
where $M_Z^2$ is the neutral-vector mass-squared matrix, $M_{\phi}^2$
is the neutral-scalar mass-squared matrix, and $M_{H_3^0}^2$ represents
the pseudoscalar mass-squared matrix. This is a very general relation. It holds
in any supersymmetric model based on an extended gauge group in which there are no 
gauge-singlet fields. Interestingly, in this model, the 
trace of the neutral-vector mass-squared matrix must include the $W_I$ fields, 
which are the neutral nondiagonal bosons of the $SU(2)_I$ group. 

For every set of values of $\lambda$, $\lambda^{\prime}$ and $m_3^2$, we 
searched numerically for the minimum of the scalar potential. 
We choose $\lambda$ and $\lambda^{\prime}$ to be as large as $1$ and 
$m_3$ to be as large as $1000$ GeV. 
If the value of $\lambda$ or $\lambda^{\prime}$ is too large, it will blow up at the 
unification scale by the renormalization group analysis as in the SM.  
Adjusting the various 
vacuum expectation values until the eigenvalues of the Higgs-boson mass
matrices are positive or zero, we read off the value of the smallest nonzero
eigenvalue of the neutral scalar mass-squared matrix. Then we vary the values of 
$\lambda$, $\lambda^{\prime}$ and $m_3^2$ to find the largest possible value 
of this smallest nonzero eigenvalue. We find that its value is about $150$ GeV.  

\section{Conclusions}
\hspace{0.2 in}We have considered the effective low-energy $SU(2)_L \times U(1)_Y \times SU(2)_I
\times U(1)_{Y^{\prime}}$ model, which can arise from the $E_6$ unification model. 
The $SU(2)_I$ is a subgroup of $SU(3)_R$ and commutes with the electric
charge operator, so the three corresponding gauge bosons are neutral.   
The gauge boson and Higgs boson sectors of the model are studied.

The extra neutral gauge bosons generally mix with each other and also with the ordinary Z boson. The electroweak 
precision data including Z-pole experiments, $m_W$ measurement and LENC experiments are used to put constraints 
on masses of extra gauge bosons and the mixings with ordinary Z bosons. 
The possible constraint from the weak charge of the proton, 
which is proposed to be measured at Jefferson Lab,
is also considered. It is found that the mixings 
are very small, namely $|\theta| \leq 0.003$. The lower mass limit for the lightest extra neutral gauge 
boson is found to be about $900$ GeV, which is somewhat higher than bounds in the literature 
mainly due to more updated data used in our analysis.

The scalar potential is highly constrained in the supersymmetric version of the model. 
An upper bound of about $150$ GeV to the mass of the lightest CP-even Higgs scalar is found. 
 
We thank John M. Finn for informing us of the proposal of the measurement of the weak charge of the proton at Jefferson Lab.
This work was supported by the National Science Foundation grant NSF-PHY-9900657.

%\newpage 
\bigskip    
\begin{appendix}
\noindent {\Large {\bf Appendix}} 
\section{Mass-squared matrix for neutral gauge bosons}
The mass-squared matrix for neutral gauge bosons is a symmetric $5 \times 5$ matrix.
\begin{eqnarray}
m_{Z}^2&=&\frac{1}{4} (g_L^2+g_Y^2) ( v_1^2+v_2^2+v_3^2), \nonumber \\
m_{12}&=&\frac{1}{4}\sqrt{g_L^2+g_Y^2} g_I (v_1^2-v_3^2), \nonumber \\
m_{13}&=&\frac{1}{12}\sqrt{g_L^2+g_Y^2} g_{Y^{\prime}} (-v_1^2 + 4 v_2^2 - v_3^2), \nonumber \\
m_{14}&=&m_{15}=\frac{1}{4}\sqrt{g_L^2+g_Y^2} g_I v_1 v_3, \nonumber \\
m_{W_I^3}^2&=&\frac{1}{4}g_I^2(v_1^2+v_3^2+n_1^2+n_2^2+{n_1^{\prime}}^2+ {n_2^{\prime}}^2), \nonumber \\
m_{23}&=&\frac{1}{12}g_I g_{Y^{\prime}} [-v_1^2+v_3^2-5(n_1^2-n_2^2+{n_1^{\prime}}^2 - {n_2^{\prime}}^2)], \nonumber \\
m_{24}&=&m_{25}=0, \nonumber \\
m_{B}^2&=&\frac{1}{36}g_{Y^{\prime}}^2 [ v_1^2 +16 v_2^2 +v_3^2+25(n_1^2+n_2^2+{n_1^{\prime}}^2+ {n_2^{\prime}}^2)], \nonumber \\
m_{34}&=&m_{35}=\frac{1}{12 \sqrt{2}}g_I g_{Y^{\prime}}[-v_1 v_3+5(n_1 n_2 +n_1^{\prime} n_2^{\prime} )], \nonumber \\
m_{44}&=&m_{55}=0, \nonumber \\
m_{W_I^{\pm}}^2&=&\frac{1}{4}g_I^2(v_1^2+v_3^2+n_1^2+n_2^2+{n_1^{\prime}}^2+ {n_2^{\prime}}^2). 
\end{eqnarray}

\section{Various Higgs boson mass-squared matrices}
The Higgs boson mass-squared matrix is obtained from
\begin{equation}
M_{ij}^2=\frac{\partial ^2 V}{\partial \phi_i \phi_j}|_{minimum} \ .
\end{equation}
\subsection {Scalar Higgs boson mass-squared matrix }
The mass-squared matrix for scalar Higgs bosons is a $7 \times 7$ symmetric matrix, $S$.
Let 
\begin{equation}
V^2 \equiv v_1^2+4v_2^2+v_3^2-5 (n_1^2+n_2^2+{n_1^{\prime}}^2+{n_2^{\prime}}^2),
\end{equation}
\begin{eqnarray}
S_{11}&=& (\lambda n_1 +\lambda ^{\prime} n_1^{\prime})^2+(\lambda ^2 + {\lambda ^{\prime}}^2) v_2^2 +\frac{1}{2} (g_L^2+g_Y^2+g_I^2+\frac{1}{9} g_{Y^{\prime}}^2) v_1^2 +\frac{1}{4} g_L^2 (v_1^2-v_2^2+v_3^2)  \nonumber \\
& & +\frac{1}{4} g_Y^2 (v_1^2-v_2^2+v_3^2) +\frac{1}{4} g_I^2 (v_1^2+v_3^2+n_2^2-n_1^2+{n_2^{\prime}}^2-{n_1^{\prime}}^2)+\frac{1}{36}g_{Y^{\prime}}^2V^2 + m_{{\cal H}}^2,  \nonumber \\
S_{12}&=& 2 (\lambda ^2 + {\lambda ^{\prime}}^2) v_1 v_2 -\frac{1}{2} (g_L^2+g_Y^2-\frac{4}{9} g_{Y^{\prime}}^2) v_1 v_2+\lambda A n_1 +\lambda^{\prime} A^{\prime} n_1^{\prime},  \nonumber \\
S_{13}&=& - (\lambda n_1 + \lambda ^{\prime} n_1^{\prime})(\lambda n_2 + \lambda ^{\prime} n_2^{\prime}) +\frac{1}{2} (g_L^2+g_Y^2+g_I^2+\frac{1}{9} g_{Y^{\prime}}^2) v_1 v_3+\frac{1}{2}g_I^2( n_1 n_2 +n_1^{\prime}  n_2^{\prime}), \nonumber \\ 
S_{14}&=&-\lambda v_3 ( \lambda n_1 + \lambda^{\prime} n_1^{\prime}) + \frac{1}{2}g_I^2 (v_1 n_2 + v_3 n_1 ) -\frac{5}{18} g_{Y^{\prime}}^2 v_1 n_2, \nonumber \\
S_{15}&=& 2 \lambda v_1 ( \lambda n_1 + \lambda^{\prime} n_1^{\prime})-\lambda v_3 ( \lambda n_2 + \lambda^{\prime} n_2^{\prime}) + \frac{1}{2}g_I^2 (v_3 n_2 - v_1 n_1 ) -\frac{5}{18} g_{Y^{\prime}}^2 v_1 n_1+\lambda A v_2, \nonumber \\
S_{16}&=&-\lambda^{\prime} v_3 ( \lambda n_1 + \lambda^{\prime} n_1^{\prime}) + \frac{1}{2}g_I^2 (v_1 n_2 ^{\prime}+ v_3 n_1^{\prime} ) -\frac{5}{18} g_{Y^{\prime}}^2 v_1 n_2^{\prime}, \nonumber \\
S_{17}&=& \lambda^{\prime} v_1 ( \lambda n_1 + \lambda^{\prime} n_1^{\prime})+\lambda^{\prime} [( \lambda n_1 + \lambda^{\prime} n_1^{\prime}) v_1- ( \lambda n_2 + \lambda^{\prime} n_2^{\prime}) v_3], \nonumber \\
& &+ \frac{1}{2}g_I^2 (v_3 n_2^{\prime} - v_1 n_1^{\prime} ) -\frac{5}{18} g_{Y^{\prime}}^2 v_1 n_1^{\prime}+\lambda^{\prime} A^{\prime} v_2, \nonumber \\
S_{22}&=& (\lambda n_1 +\lambda ^{\prime} n_1^{\prime})^2+(\lambda n_2 + {\lambda ^{\prime}} n_2^{\prime})^2+ (\lambda ^2 + {\lambda ^{\prime}} ^2)(v_1^2+v_3^2) +\frac{1}{2} (g_L^2+g_Y^2+\frac{16}{9} g_{Y^{\prime}}^2) v_2^2  \nonumber \\
& & -\frac{1}{4} g_L^2 (v_1^2-v_2^2+v_3^2) +\frac{1}{4} g_Y^2 (v_1^2-v_2^2+v_3^2)+\frac{1}{9}g_{Y^{\prime}}^2V^2 + m_{{\sl H_2}}^2 , \nonumber \\
S_{23}&=& 2 (\lambda ^2 + {\lambda ^{\prime}}^2) v_2 v_3 -\frac{1}{2} (g_L^2+g_Y^2-\frac{4}{9} g_{Y^{\prime}}^2) v_2 v_3-\lambda A n_2 -\lambda^{\prime} A^{\prime} n_2^{\prime},  \nonumber \\
S_{24}&=&2 \lambda v_2 ( \lambda n_2 + \lambda^{\prime} n_2^{\prime}) -\frac{10}{9} g_{Y^{\prime}}^2 v_2 n_2 -\lambda A v_3, \nonumber \\
S_{25}&=&2 \lambda v_2 ( \lambda n_1 + \lambda^{\prime} n_1^{\prime}) -\frac{10}{9} g_{Y^{\prime}}^2 v_2 n_1 +\lambda A v_1, \nonumber \\
S_{26}&=&2 \lambda^{\prime} v_2 ( \lambda n_2 + \lambda^{\prime} n_2^{\prime}) -\frac{10}{9} g_{Y^{\prime}}^2 v_2 n_2 ^{\prime}-\lambda^{\prime} A^{\prime} v_3, \nonumber \\
S_{27}&=&2 \lambda^{\prime} v_2 ( \lambda n_1 + \lambda^{\prime} n_1^{\prime}) -\frac{10}{9} g_{Y^{\prime}}^2 v_2 n_1^{\prime} +\lambda^{\prime} A^{\prime} v_1, \nonumber \\
S_{33}&=& (\lambda n_2+\lambda ^{\prime} n_2^{\prime})^2+(\lambda ^2 + {\lambda ^{\prime}}^2) v_2^2 +\frac{1}{2} (g_L^2+g_Y^2+g_I^2+\frac{1}{9} g_{Y^{\prime}}^2) v_3^2 +\frac{1}{4} g_L^2 (v_1^2-v_2^2+v_3^2)  \nonumber \\
& & +\frac{1}{4} g_Y^2 (v_1^2-v_2^2+v_3^2) +\frac{1}{4} g_I^2 (v_1^2+v_3^2+n_1^2-n_2^2+{n_1^{\prime}}^2-{n_2^{\prime}}^2) +\frac{1}{36}g_{Y^{\prime}}^2V^2 + m_{{\cal H}}^2,  \nonumber \\
S_{34}&=&  2 (\lambda n_2 + {\lambda ^{\prime}} n_2^{\prime}) \lambda v_3 - (\lambda n_1 + {\lambda ^{\prime}} n_1^{\prime}) \lambda v_1+\frac{1}{2} g_I^2 ( v_1 n_1-v_3 n_2)-\frac{5}{18}g_{Y^{\prime}}^2 v_3 n_2-\lambda A v_2,  \nonumber \\
S_{35}&=& - (\lambda n_2 + {\lambda ^{\prime}} n_2^{\prime}) \lambda v_1+\frac{1}{2} g_I^2 ( v_1 n_2+v_3 n_1)-\frac{5}{18}g_{Y^{\prime}}^2 v_3 n_1,  \nonumber \\
S_{36}&=&  2 (\lambda n_2 + {\lambda ^{\prime}} n_2^{\prime}) \lambda^{\prime} v_3 - (\lambda n_1 + {\lambda ^{\prime}} n_1^{\prime}) \lambda^{\prime} v_1+\frac{1}{2} g_I^2 ( v_1 n_1^{\prime}-v_3 n_2^{\prime})-\frac{5}{18}g_{Y^{\prime}}^2 v_3 n_2^{\prime}-\lambda^{\prime} A^{\prime} v_2,  \nonumber \\
S_{37}&=& - (\lambda n_2 + {\lambda ^{\prime}} n_2^{\prime}) \lambda^{\prime} v_1+\frac{1}{2} g_I^2 ( v_1 n_2^{\prime}+v_3 n_1^{\prime})-\frac{5}{18}g_{Y^{\prime}}^2 v_3 n_1^{\prime}, \nonumber \\
S_{44}&=& \lambda^2( v_2^2+v_3^2) +\frac{1}{2} (g_I^2+\frac{25}{9} g_{Y^{\prime}}^2) n_2^2 +\frac{1}{4}g_I^2 (v_1^2-v_3^2+n_1^2+n_2^2-{n_1^{\prime}}^2+{n_2^{\prime}}^2) \nonumber \\
      & & -\frac{5}{36}g_{Y^{\prime}}^2 V^2 + m_{{\sl N}}^2,  \nonumber \\
S_{45}&=&-\lambda ^2 v_1 v_3 +\frac{1}{2} g_I^2 (v_1 v_3+n_1 n_2+n_1^{\prime} n_2^{\prime})+\frac{25}{18} g_{Y^{\prime}}^2 n_1 n_2, \nonumber \\ 
S_{46}&=&\lambda \lambda^{\prime}( v_2^2+ v_3^2) +\frac{1}{2} g_I^2 (n_1 n_1^{\prime} +n_2 n_2^{\prime})+\frac{25}{18} g_{Y^{\prime}}^2 n_2 n_2^{\prime} -m_3^2, \nonumber \\ 
S_{47}&=&-\lambda \lambda^{\prime} v_1 v_3 +\frac{1}{2} g_I^2 (n_1 n_2^{\prime} -n_2 n_1^{\prime})+\frac{25}{18} g_{Y^{\prime}}^2 n_2 n_1^{\prime}, \nonumber \\ 
S_{55}&=& \lambda^2( v_1^2+v_2^2) +\frac{1}{2} (g_I^2+\frac{25}{9} g_{Y^{\prime}}^2) n_1^2 +\frac{1}{4} g_I^2 (v_3^2-v_1^2+n_1^2+n_2^2+{n_1^{\prime}}^2-{n_2^{\prime}}^2) \nonumber \\ 
& &-\frac{5}{36}g_{Y^{\prime}}^2V^2 + m_{{\sl N}}^2,  \nonumber \\
S_{56}&=&-\lambda ^2 v_1 v_3 +\frac{1}{2} g_I^2 (n_2 n_1^{\prime} -n_1 n_2^{\prime})+\frac{25}{18} g_{Y^{\prime}}^2 n_1 n_2^{\prime}, \nonumber \\ 
S_{57}&=&\lambda \lambda^{\prime}( v_1^2+ v_2^2) +\frac{1}{2} g_I^2 (n_1 n_1^{\prime} +n_2 n_2^{\prime})+\frac{25}{18} g_{Y^{\prime}}^2 n_1 n_1^{\prime} -m_3^2, \nonumber \\ 
S_{66}&=& {\lambda^{\prime}}^2( v_2^2+v_3^2) +\frac{1}{2} (g_I^2+\frac{25}{9} g_{Y^{\prime}}^2) {n_2^{\prime}}^2 +\frac{1}{4} g_I^2 (v_1^2-v_3^2+{n_1^{\prime}}^2+{n_2^{\prime}}^2-n_1^2+n_2^2) \nonumber \\ 
& &-\frac{5}{36}g_{Y^{\prime}}^2V^2 + m_{{\sl {N^{\prime}}}}^2,  \nonumber \\
S_{67}&=&- {\lambda^{\prime}}^2 v_1 v_3 +\frac{1}{2} g_I^2 (v_1 v_3 +n_1 n_2+n_1^{\prime} n_2^{\prime})+\frac{25}{18} g_{Y^{\prime}}^2 n_1^{\prime} n_2^{\prime}, \nonumber \\ 
S_{77}&=& {\lambda^{\prime}}^2( v_1^2+v_2^2) +\frac{1}{2} (g_I^2+\frac{25}{9} g_{Y^{\prime}}^2) {n_1^{\prime}}^2 +\frac{1}{4} g_I^2 (v_3^2-v_1^2+{n_1^{\prime}}^2+{n_2^{\prime}}^2+n_1^2-n_2^2) \nonumber \\ 
& &-\frac{5}{36}g_{Y^{\prime}}^2V^2 + m_{{\sl {N^{\prime}}}}^2.
\end{eqnarray}
\subsection {Pseudocalar Higgs boson mass-squared matrix }
The mass-squared matrix for pseudoscalar Higgs bosons is also a $7 \times 7$ symmetric matrix, $P$. 
\begin{eqnarray}
P_{11}&=& (\lambda n_1 +\lambda ^{\prime} n_1^{\prime})^2+(\lambda ^2 + {\lambda ^{\prime}}^2) v_2^2 +\frac{1}{4} g_L^2 (v_1^2-v_2^2+v_3^2) +\frac{1}{4} g_Y^2 (v_1^2-v_2^2+v_3^2) \nonumber \\
& &+\frac{1}{4} g_I^2 (v_1^2+v_3^2+n_2^2-n_1^2+{n_2^{\prime}}^2-{n_1^{\prime}}^2)+\frac{1}{36}g_{Y^{\prime}}^2V^2 + m_{{\cal H}}^2,  \nonumber \\
P_{12}&=& \lambda A n_1 +\lambda^{\prime} A^{\prime} n_1^{\prime},  \nonumber \\
P_{13}&=& - (\lambda n_1 + \lambda ^{\prime} n_1^{\prime})(\lambda n_2 + \lambda ^{\prime} n_2^{\prime}) +\frac{1}{2}g_I^2( n_1 n_2 +n_1^{\prime}  n_2^{\prime}), \nonumber \\ 
P_{14}&=&\lambda v_3 ( \lambda n_1 + \lambda^{\prime} n_1^{\prime}) - \frac{1}{2}g_I^2  v_3 n_1,  \nonumber \\
P_{15}&=&-\lambda v_3 ( \lambda n_2 + \lambda^{\prime} n_2^{\prime}) + \frac{1}{2}g_I^2 v_3 n_2 +\lambda A v_2, \nonumber \\
P_{16}&=&\lambda^{\prime} v_3 ( \lambda n_1 + \lambda^{\prime} n_1^{\prime}) - \frac{1}{2}g_I^2 v_3 n_1^{\prime},  \nonumber \\
P_{17}&=& - \lambda^{\prime}v_3 ( \lambda n_2 + \lambda^{\prime} n_2^{\prime}) + \frac{1}{2}g_I^2 v_3 n_2^{\prime}+\lambda^{\prime} A^{\prime} v_2, \nonumber \\
P_{22}&=& (\lambda n_1 +\lambda ^{\prime} n_1^{\prime})^2+(\lambda n_2 + {\lambda ^{\prime}} n_2^{\prime})^2+ (\lambda ^2 + {\lambda ^{\prime}} ^2)(v_1^2+v_3^2 ) -\frac{1}{4} g_L^2 (v_1^2-v_2^2+v_3^2)  \nonumber \\
& &+\frac{1}{4}g_Y^2(v_1^2-v_2^2+v_3^2)  +\frac{1}{9}g_{Y^{\prime}}^2V^2 + m_{{\sl H_2}}^2,  \nonumber \\
P_{23}&=& -\lambda A n_2 -\lambda^{\prime} A^{\prime} n_2^{\prime},  \nonumber \\
P_{24}&=&\lambda A v_3, \nonumber \\
P_{25}&=&-\lambda A v_1, \nonumber \\
P_{26}&=&\lambda^{\prime} A^{\prime} v_3, \nonumber \\
P_{27}&=&-\lambda^{\prime} A^{\prime} v_1, \nonumber \\
P_{33}&=& (\lambda n_2+\lambda ^{\prime} n_2^{\prime})^2+(\lambda ^2 + {\lambda ^{\prime}}^2) v_2^2+\frac{1}{4} g_L^2 (v_1^2-v_2^2+v_3^2) +\frac{1}{4} g_Y^2 (v_1^2-v_2^2+v_3^2) \nonumber \\
& & +\frac{1}{4} g_I^2 (v_1^2+v_3^2+n_1^2-n_2^2+{n_1^{\prime}}^2-{n_2^{\prime}}^2) +\frac{1}{36}g_{Y^{\prime}}^2V^2 + m_{{\cal H}}^2, \nonumber \\
P_{34}&=&   - (\lambda n_1 + {\lambda ^{\prime}} n_1^{\prime}) \lambda v_1+\frac{1}{2} g_I^2  v_1 n_1-\lambda A v_2,  \nonumber \\
P_{35}&=&  (\lambda n_2 + {\lambda ^{\prime}} n_2^{\prime}) \lambda v_1-\frac{1}{2} g_I^2  v_1 n_2, \nonumber \\
P_{36}&=& - (\lambda n_1 + {\lambda ^{\prime}} n_1^{\prime}) \lambda^{\prime} v_1+\frac{1}{2} g_I^2 v_1 n_1^{\prime}-\lambda^{\prime} A^{\prime} v_2,  \nonumber \\
P_{37}&=& (\lambda n_2 + {\lambda ^{\prime}} n_2^{\prime}) \lambda^{\prime} v_1-\frac{1}{2} g_I^2 v_1 n_2^{\prime}, \nonumber \\
P_{44}&=& \lambda^2( v_2^2+v_3^2) +\frac{1}{4} g_I^2 (v_1^2-v_3^2+n_1^2+n_2^2-{n_1^{\prime}}^2+{n_2^{\prime}}^2) -\frac{5}{36}g_{Y^{\prime}}^2V^2 + m_{{\sl N}}^2,  \nonumber \\
P_{45}&=&-\lambda ^2 v_1 v_3 +\frac{1}{2} g_I^2 (v_1 v_3+n_1^{\prime} n_2^{\prime}), \nonumber \\ 
P_{46}&=&\lambda \lambda^{\prime}( v_2^2+ v_3^2) +\frac{1}{2} g_I^2 n_1 n_1^{\prime}  -m_3^2, \nonumber \\ 
P_{47}&=&-\lambda \lambda^{\prime} v_1 v_3 -\frac{1}{2} g_I^2 n_1 n_2^{\prime},  \nonumber \\ 
P_{55}&=& \lambda^2( v_1^2+v_2^2) +\frac{1}{4} g_I^2 (v_3^2-v_1^2+n_1^2+n_2^2+{n_1^{\prime}}^2-{n_2^{\prime}}^2) -\frac{5}{36}g_{Y^{\prime}}^2V^2 + m_{{\sl N}}^2,  \nonumber \\
P_{56}&=&-\lambda ^2 v_1 v_3 -\frac{1}{2} g_I^2 n_2 n_1^{\prime}, \nonumber \\ 
P_{57}&=&\lambda \lambda^{\prime}( v_1^2+ v_2^2) +\frac{1}{2} g_I^2 n_2 n_2^{\prime} -m_3^2, \nonumber \\ 
P_{66}&=& {\lambda^{\prime}}^2( v_2^2+v_3^2) +\frac{1}{4} g_I^2 (v_1^2-v_3^2+{n_1^{\prime}}^2+{n_2^{\prime}}^2-n_1^2+n_2^2) -\frac{5}{36}g_{Y^{\prime}}^2V^2 + m_{{\sl {N^{\prime}}}}^2,  \nonumber \\
P_{67}&=&- {\lambda^{\prime}}^2 v_1 v_3 +\frac{1}{2} g_I^2 (v_1 v_3 +n_1 n_2), \nonumber \\ 
P_{77}&=& {\lambda^{\prime}}^2( v_1^2+v_2^2) +\frac{1}{4} g_I^2 (v_3^2-v_1^2+{n_1^{\prime}}^2+{n_2^{\prime}}^2+n_1^2-n_2^2) -\frac{5}{36}g_{Y^{\prime}}^2V^2 + m_{{\sl {N^{\prime}}}}^2.
\end{eqnarray}
\subsection {Charged Higgs boson mass-squared matrix }
The mass-squared matrix for charged Higgs bosons is a $3 \times 3$ symmetric matrix, $C$.
\begin{eqnarray}
C_{11}&=& (\lambda n_1 +\lambda ^{\prime} n_1^{\prime})^2+\frac{1}{4} g_L^2 (v_1^2-v_2^2+v_3^2) +\frac{1}{4} g_I^2 (v_1^2+v_3^2+n_2^2-n_1^2+{n_2^{\prime}}^2-{n_1^{\prime}}^2) \nonumber \\
& &+ \frac{1}{4} g_Y^2 (v_1^2-v_2^2+v_3^2) +\frac{1}{36}g_{Y^{\prime}}^2V^2 + m_{{\cal H}}^2,  \nonumber \\
C_{12}&=&  (\lambda ^2 + {\lambda ^{\prime}}^2) v_1 v_2 -\frac{1}{2} g_L^2 v_1 v_2+\lambda A n_1 +\lambda^{\prime} A^{\prime} n_1^{\prime},  \nonumber \\
C_{13}&=& - (\lambda n_1 + \lambda ^{\prime} n_1^{\prime})(\lambda n_2 + \lambda ^{\prime} n_2^{\prime}) +\frac{1}{2} g_L^2 v_1 v_3+\frac{1}{2}g_I^2( v_1 v_3 n_1 n_2 +n_1^{\prime}  n_2^{\prime}), \nonumber \\ 
C_{22}&=& (\lambda n_1 +\lambda ^{\prime} n_1^{\prime})^2+(\lambda n_2 + {\lambda ^{\prime}} n_2^{\prime})^2+ \frac{1}{4} g_L^2 (v_1^2+v_2^2+v_3^2),  \nonumber \\
& &-\frac{1}{4}g_Y^2(v_1^2-v_2^2+v_3^2)  +\frac{1}{9}g_{Y^{\prime}}^2V^2 + m_{{\sl H_2}}^2, \nonumber \\
C_{23}&=&  (\lambda ^2 + {\lambda ^{\prime}}^2) v_2 v_3 -\frac{1}{2} g_L^2 v_2 v_3-\lambda A n_2 -\lambda^{\prime} A^{\prime} n_2^{\prime},  \nonumber \\
C_{33}&=& (\lambda n_2 +\lambda ^{\prime} n_2^{\prime})^2+\frac{1}{4} g_L^2 (-v_1^2+v_2^2+v_3^2) +\frac{1}{4} g_I^2 (v_1^2+v_3^2+n_1^2-n_2^2+{n_1^{\prime}}^2-{n_2^{\prime}}^2) \nonumber \\
& &+\frac{1}{4} g_Y^2 (v_1^2-v_2^2+v_3^2) +\frac{1}{36}g_{Y^{\prime}}^2V^2 + m_{{\cal H}}^2. 
\end{eqnarray}

\end{appendix}

\bibliographystyle{unsrt}

\end{document}